\documentclass[prd,superscriptaddress,twocolumn,nopreprintnumbers,floatfix,nofootinbib]{revtex4}

\usepackage{amssymb}
\usepackage{amsmath}
\usepackage{graphicx}
\usepackage{dcolumn}
\usepackage{hyperref}
\usepackage{color,units}
\usepackage{lineno}
\usepackage{xspace}
\usepackage{mathtools}
\usepackage{physics}
\usepackage{tensor}
\usepackage[dvipsnames]{xcolor}
\usepackage[normalem]{ulem}

\newcommand{\bilby}{{\sc Bilby}\xspace}

\newcommand{\lalsuite}{{\sc LALSuite}\xspace}

\newcommand{\M}{\mathcal{M}}

\newcommand{\SPA}{School of Physics and Astronomy, Monash University, Vic 3800, Australia}
\newcommand{\OzGravMonash}{OzGrav: The ARC Centre of Excellence for Gravitational Wave Discovery, Clayton VIC 3800, Australia}

\begin{document}

\title{Measuring the neutron star equation of state with gravitational waves: \\
the first forty binary neutron star mergers}

\author{Francisco Hernandez Vivanco}
\email{francisco.hernandezvivanco@monash.edu}
\affiliation{\SPA}
\affiliation{\OzGravMonash}

\author{Rory Smith}
\email{rory.smith@monash.edu}
\author{Eric Thrane}
\author{Paul D. Lasky}
\author{Colm Talbot}
\affiliation{\SPA}
\affiliation{\OzGravMonash}

\author{Vivien Raymond}
\affiliation{School of Physics and Astronomy, Cardiff University, Cardiff, CF24 3AA, United Kingdom}

\date{\today}

\begin{abstract}
Gravitational waves from binary neutron star coalescences contain rich information about matter at supranuclear densities encoded by the neutron star equation of state.
We can measure the equation of state by analyzing the tidal interactions between neutron stars, which is quantified by the tidal deformability.
Multiple merger events are required to probe the equation of state over a range of neutron star masses.
The more events included in the analysis, the stronger the constraints on the equation of state.
In this paper, we build on previous work to explore the constraints that LIGO and Virgo are likely to place on the neutron star equation of state by combining the first forty binary neutron star detections, a milestone we project to be reached during the first year of accumulated design-sensitivity data.
We carry out Bayesian inference on a realistic mock dataset of binaries to obtain posterior distributions for neutron star tidal parameters. 
In order to combine posterior samples from multiple observations, we employ a random forest regressor, which allows us to efficiently interpolate the likelihood distribution.
Assuming a merger rate of \unit[1540]{Gpc$^{-3}$ yr$^{-1}$} and a LIGO-Virgo detector network operating for one year at the sensitivity of the third-observation run, plus an additional eight months of design sensitivity, we find that the radius of a $\unit[1.4]{M_\odot}$ neutron star can be constrained to better than \unit[$\sim 10$]{\%} at 90\% confidence.
The pressure at twice the nuclear saturation density can be constrained to better than \unit[$\sim 45$]{\%} at 90\% confidence. We show that these results are robust to our choice of parametrization for the neutron star equation of state. Finally, we add an appendix following publication of the paper in the journal, showing the posterior distribution of the maximum neutron star mass allowed by the equation of state. We find that the maximum mass can be constrained to \unit[$\sim 0.3$]{$M_\odot$} at 90\% confidence.
\end{abstract}

\maketitle
\section{Introduction}
With the improving sensitivity of LIGO~\cite{Aasi:2015,aligo:2015} and Virgo~\cite{Acernese:2014}, we expect to increase the number of binary neutron star detections using gravitational waves~\cite{Abbott2018}.
Binary neutron star coalescences provide rich information about the neutron star equation of state, which can be constrained by measuring the neutron star's tidal interactions.
These tidal interactions are determined by the neutron star's dimensionless tidal deformability $\Lambda$, which determines how the neutron star's quadruple moment changes in response to the companion's tidal field (e.g. \cite{Flanagan:2008,Hinderer:2010}).
The gravitational-wave detection of a binary neutron star coalescence, GW170817~\cite{LSC_GW170817}, placed the first constraints on $\Lambda$, ruling out some of the stiffest proposed equations of state~\cite{LSC_EoS:2018}. 

In order to obtain a better measurement of the neutron star equation of state, it is necessary to combine information from an ensemble of binary neutron stars. 
The first fully Bayesian study that combined information to constrain the equation of state using tidal interactions was proposed by~\citet{DelPozzo:2013}. They simulated an astrophysically distributed gravitational-wave dataset and showed that a few tens of observations were enough to constrain the tidal deformability $\lambda=\Lambda/G(Gm/c^2)^5$ to about 10\% at a reference mass of \unit[1.4]{$M_\odot$}.
The method in Ref.~\cite{DelPozzo:2013} fits the tidal deformability $\lambda$ by Taylor expanding $\lambda(m)$ around a fiducial mass of $m=1.4M_\odot$. Then, Lackey and Wade~\cite{Lackey:2015} showed that $\lambda$ can be measured at \unit[10-50]{\%} in the range between \unit[1-2]{$M_\odot$}, where most of the information comes from the loudest $\sim 5$ events. Their method used Markov chain Monte Carlo simulations to fit a model that follows the piecewise polytrope parametrization of the equation of state.  In order to reduce the computational cost of their study, they used Fisher matrix approximations in some of the events contained in their mock data study. 
\citet{Agathos:2015} improved on~\cite{Lackey:2015} by including waveforms with tidal effects up to 1 post-Newtonian (1PN) order, neutron-star spins, and termination of the waveform at the contact frequency.
While the results from~\citet{Agathos:2015} are broadly consistent with the findings of~\cite{Lackey:2015}, Agathos et al.\ find that a poorly chosen neutron-star mass prior can bias inferences about the equation of state.

In this paper, we study how well we can constrain the equation of state with gravitational-wave observations using the first forty expected binary neutron star detections from the LIGO-Virgo network. We perform a realistic mock data study, which introduces a few innovations: (1) we use a reduced order quadrature model waveform based on the {\sc IMRPhenomD\_NRTidal} approximant~\cite{Husa:2016,Dietrich:2017,Khan:2016} which significantly reduces the cost of parameter estimation; (2) we run full parameter estimation on all the events in our dataset that contain significant information about the equation of state and (3) we introduce a new interpolation method for combining events, yielding an interpolation error of $\approx 0.3\%$.

The problem of combining equation of state constraints from multiple events is an example of an interesting general class of problems in which (1) the prior for an individual measurement is conditional on some hyper-parameter; (2) the conditional prior spans some $d$-dimensional subspace with fewer dimensions than the full $D$-dimensional parameter space and (3) the equation of state curves are deterministic, i.e., they live exactly on a curve rather than have a probability defined over the whole parameter space.
As we discuss in greater detail below, standard techniques from hierarchical modelling fail when applied to such problems.
In this paper, we discuss the subtleties associated with combining events.
We document the challenges arising from this unique class of problem and propose a new solution, which employs a random forest regressor to interpolate the likelihood distribution.

This paper is organized as follows.
In Section~\ref{sec:Method}, we describe the difficulties of combining a population of gravitational-wave measurements to constrain the equation of state. In order to solve this problem, we describe a new method that is based on interpolating the likelihood distribution using a random forest regressor.
In Section~\ref{sec:Results}, we present constraints that the LIGO-Virgo detector network will place on the equation of state after the first forty gravitational-wave observations.
We discuss our results in Section~\ref{sec:Discussion} and conclude in Section~\ref{sec:Conclusion}.

\section{Method}\label{sec:Method}
\subsection{Equation of state parametrization}
Gravitational-wave observations can be used to constrain the equation of state by measuring the tidal deformability of neutron stars. If we can measure these tidal interactions, we can place constraints on relations that model the equation of state such as pressure $p$ and density $\rho$ (see e.g.~\cite{Ozel_review:2016}). While we cannot measure pressure and density directly, we can map $p(\rho)$ onto the measurable astrophysical parameters $\Lambda(m)$ using the Tolman-Oppenheimer-Volkoff equations and the relationship between the tidal deformability, second Love number, and radius~\cite{Lindblom:1992, Hinderer:2008}. 

\citet{Read:2009} showed that $p(\rho)$ can be modeled assuming a piecewise polytropic relation between pressure and density, where each polytrope is defined between three different sections, $i=1,2,3$. Each polytrope is modeled by
\begin{align}
    p=K_i\rho^{\Gamma_i}.
\end{align}
Here, $K_i$ is a constant and $\Gamma_i$ is the adiabatic index. The density transitions occur at $\rho_1=\unit[10^{14.7}]{g \; cm^{-3}}$ and $\rho_2=\unit[10^{15}]{g \; cm^{-3}}$~\cite{Read:2009,Flanagan:2008,Lattimer:2006xb,Hinderer:2008}. 
Demanding that $p(\rho)$ is continuous implies the equation of state can be fully determined with only four numbers: $\Gamma_i$ for $i =1,2,3$ and $p_1$, the value of pressure at $\rho_1$. 

The mapping of $p(\rho)$ onto $\Lambda(m)$ allows us to cast the neutron star equation of state measurement as a hierarchical model, where the neutron star parameters $\Lambda$ and $m$ are conditional on “hyper-parameters” $\Upsilon$, which describe the shape of the piecewise polytrope. 

In this study, we follow the piecewise polytrope model~\cite{Read:2009} to sample over the equation of state hyper-parameters. We choose the hyper-priors following the discussions from~\cite{Carney:2018,Lackey:2015}. The hyper-priors are shown in Tab.~\ref{tab:hyper_priors}. Additionally, when we sample over the piecewise polytrope hyper-parameters, we impose three astrophysically-motivated constraints: 
\begin{enumerate}
    \item The maximum speed of sound cannot exceed the speed of light. However, since the piecewise polytrope only fits the equations of state to an accuracy of $\sim 4$ \%, we set the causality constraint only when the speed of sound exceeds 1.1 times the speed of light.
    \item We only allow equations of state with maximum masses above $m=1.97$ $M_\odot$, which is consistent with the maximum TOV mass observed from pulsar PSR J0348+0432 ~\cite{Antoniadis1233232} and PSR J0740+6620~\cite{Cromartie:2019}.
    \item We restrict the tidal deformability to $\Lambda \leq 5000$, conservatively consistent with the gravitational-wave measurement GW170817~\cite{LSC_GW170817}. 
\end{enumerate}

\begin{table}[t!]
    \begin{ruledtabular}
    \begin{tabular}{cccc}
    Hyper-parameter                   & Prior     & Minimum & Maximum \\
    \hline
    $\log (p_1 / [\text{dyne/cm}^2])$ & Uniform   & 33.6    & 34.8   \\
    $\Gamma_1$                        & Uniform   & 2       & 4.5       \\
    $\Gamma_2$                        & Uniform   & 1.1     & 4.5    \\
    $\Gamma_3$                        & Uniform   & 1.1     & 4.5    \\ 
    \end{tabular}
    \end{ruledtabular}
    \caption{
    Hyper-prior distributions used in our study. We model the equation of state with three polytropes and assume that we know the equation of state below the pressure $p_1$. The hyper-priors are chosen based on the discussions from Refs.~\cite{Carney:2018,Lackey:2015} and are set to cover a wide range of equations of state.
    }
    \label{tab:hyper_priors}
\end{table}

\subsection{Formalism}\label{subsec:Formalism}
Our objective is to calculate the posterior distribution on the hyper-parameters $\Upsilon=\{ \Gamma_1, \Gamma_2, \Gamma_3, p_1 \}$ that describe the equation of state. The posterior is given by Bayes theorem,
\begin{align}
    p(\Upsilon | \vec{d})
    = \frac{{\cal L}(\vec{d}|\Upsilon)\pi(\Upsilon)}{\mathcal{Z}_\Upsilon} .
\end{align}
Here, we introduce the likelihood ${\cal L}(\vec{d}|\Upsilon)$, the hyper-evidence $\mathcal{Z}_\Upsilon$ and the hyper-prior $\pi (\Upsilon)$. The hyper-prior $\pi (\Upsilon)$, in our case, describes our prior belief about the hyper-parameters $\Upsilon$ that model the equation of state. 

\begin{figure}[!t]
    \centering
    \includegraphics[width=\columnwidth]{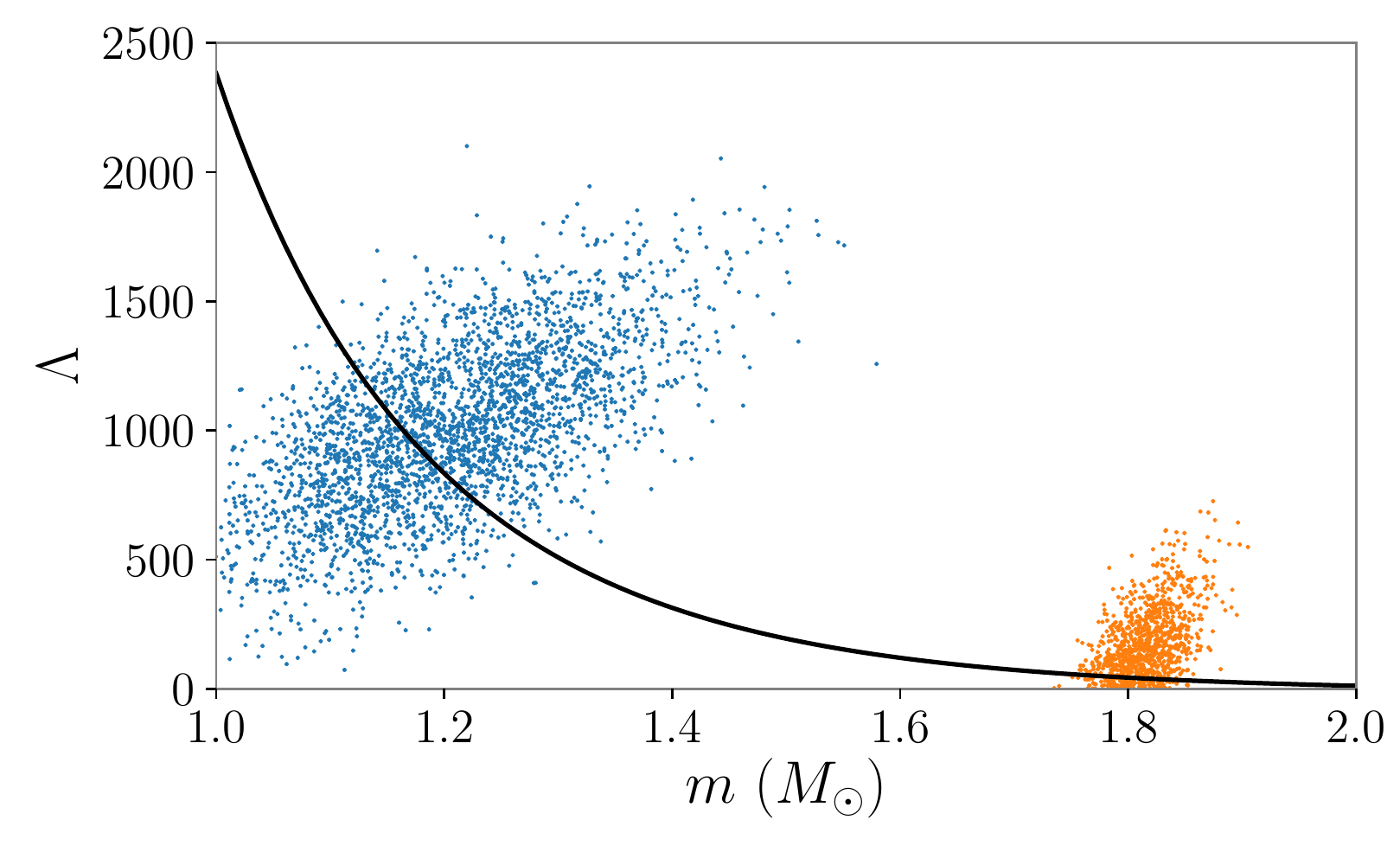}
    \caption{
     Example of posterior samples from parameter estimation of two binary neutron star observations. 
     Each gravitational-wave measurement results in posterior samples on the $\Lambda(m)$ curve, represented by blue and orange dots.  
     The equation of state is defined by an infinitesimally thin curve, represented by the black solid curve. None of the posterior samples fall exactly on the black curve. This is the reason why the usual hierarchical Bayesian inference is not directly applicable to this problem.  
    }
    \label{fig:undersampling}
\end{figure}
We can think of this problem as follows.
We have some total likelihood function, which is the product of $N$ single-event likelihoods
\begin{align} 
    \label{eq:likelihood_product}
    {\cal L}_\text{tot}(\vec{d}|\Upsilon)
    = \prod_i^N \int d\theta_i
    {\cal L}(d_i|\theta_i)
    \pi(\theta_i | \Upsilon) . 
\end{align}
Here, $\theta_i$ are the waveform parameters for the $i$th event, ${\cal L}(d_i|\theta_i)$ is the likelihood function assuming the signal hypothesis and $\pi(\theta_i|\Upsilon)$ is the conditional prior for parameters $\theta_i$ given hyper-parameters $\Upsilon$.
In our specific problem, $\theta_i$ refers to the 17 binary neutron star parameters and the hyper-parameters in $\Upsilon$ depend on the parameters that determine the equation of state, which we denote $\omega=(m,\Lambda)$.
We introduce $\kappa$ to denote the parameters in $\theta$ that are not in $\omega$.
Equation~\ref{eq:likelihood_product} is challenging to compute. We would normally analyze each event individually and then ``recycle'' the posterior samples (see e.g.~\cite{Thrane:2018qnx}):
\begin{align}
    {\cal L}_\text{tot}(\vec{d}|\Upsilon) = &
    \prod_i^N \frac{\mathcal{Z}_{\O}^i}{n_i} 
    \sum_k^{n_i} \frac{\pi(\theta_i^k|\Upsilon)}{\pi(\theta_i^k|\O)} \\
     = &
    \prod_i^N \frac{\mathcal{Z}_{\O}^i}{n_i} 
    \sum_k^{n_i} \frac{\pi(\omega_i^k|\Upsilon)}{\pi(\omega_i^k|\O)}.
\end{align}
Here, we sum over $n_i$ posterior samples, where $\pi(\theta_i^k|\O)$ is the fiducial prior used for the initial analysis and $\mathcal{Z}_{\O}$ is the corresponding signal evidence.
Since $\pi(\theta_i^k|\omega)$ is described by a curve in the $(\Lambda,m)$ plane, no posterior sample will fall exactly on this curve (see Fig.~\ref{fig:undersampling}), and so the recycling procedure fails because $\pi(\theta_i^k|\Upsilon)$ always evaluates to zero.
In the parlance of importance sampling, our proposal distribution fails to provide samples for our target distribution.

In order to solve this problem, we devise a method to interpolate the likelihood marginalized over nuisance parameters.
The method is as follows.
\begin{enumerate}
    \item Run parameter estimation on each event using the fiducial prior $\pi(\theta|\O)$.
    The priors used in our injection set are shown in Tab.~\ref{tab:priors}. For simplicity, we consider non-spinning binary neutron stars.
    \item The fiducial parameter estimation yields samples from the posterior distribution in $\omega$.
    The density of these samples is proportional to the height of the marginalized likelihood function ${\cal L}_\kappa(d|\omega)$.
    Using the posterior sample density as a rough guide, we chose ``interpolation points'' $\{\omega_j\}$ for which we calculate the $\kappa$-marginalized likelihood in the next step:
    \begin{align}
        {\cal L}_\kappa(d|\omega) \equiv
        \int d\kappa \, {\cal L}(d|\omega,\kappa) \pi(\kappa).
    \end{align}
    The values of $\{\omega_j\}$ are selected to give us the best possible representation of the marginalized likelihood so we can interpolate between different values of $\{\omega_j\}$.
    Additionally, we draw random samples from the prior to better sample the tails of the distribution.
    \item We launch a new inference job for each interpolation point $\{\omega_j\}$. These inference jobs are identical to the original inference job except we employ a delta-function prior on $\omega$ so that these parameters are fixed to the values of the interpolation point.
    All of these inference jobs can be run in parallel so this step takes less time than the previous step, though it requires more compute cores.
    Since the metric perturbation depends only on the intrinsic parameters, it can be calculated just once at the beginning of the job and cached for later use.
    Calculating the strain given the metric perturbation is an extremely fast calculation, enabling these new jobs to run comparatively quickly.
   The Bayesian evidence given by each job represents a marginalized likelihood ${\cal L}_\kappa(d|\omega)$.
    \item We create a function which interpolates between interpolation points ${\cal L}_\kappa^\text{int}(d|\omega)$ with a random forest regressor~\cite{Breiman:2001}. Once the random forest is trained, we can give it a new value of $\omega\notin\{\omega_j\}$, and the random forest provides an estimate of the marginalized likelihood ${\cal L}_\kappa(d|\omega)$.
    \item Using our new interpolated marginal likelihood, the total likelihood for $N$ events is
    \begin{align}
        \label{eq:total_interpolated_likelihood}
        {\cal L}_\text{tot}(d|\Upsilon) = & \prod_i^N \int d\omega_i
        {\cal L}_\kappa^\text{int}(d_i|\omega_i)
        \pi(\omega_i|\Upsilon) .
    \end{align}
    As we change $\Upsilon$, we obtain different curves in $\omega$, which do not pass through any of our original samples, but we are able to evaluate the hyper-likelihood anyway because we are able to interpolate throughout $\omega$.
    In practice, the integral in Eq.~\ref{eq:total_interpolated_likelihood} is carried out on a discrete grid so that the integral is replaced by a sum over mass bins.
        \begin{align}
        {\cal L}_\text{tot}(d|\Upsilon) = & \prod_i^N \int d\omega_i
        {\cal L}_\kappa^\text{int}(d_i|\omega_i)
        \pi(\omega_i|\Upsilon) \\
        {\cal L}_\text{tot}(d|\Upsilon) = & \prod_i^N \int d\omega_i
        {\cal L}_\kappa^\text{int}(d_i|\omega_i)
        \pi(m_1^{i}, m_2^{i})
        \nonumber\\
        & \delta\Big(\Lambda_1 - \Lambda'_1(m_1|\Upsilon)\Big)
        \delta\Big(\Lambda_2 - \Lambda'_2(m_2|\Upsilon)\Big)
        \nonumber\\
        = & \prod_i^N \int dm^i_1 \int dm^i_2 \,
        \pi(m_1^{i}, m_2^{i}) \nonumber\\
        & {\cal L}_\kappa^\text{int}\bigg(d_i\bigg|m^i_1, m^i_2, \Lambda_1^i(m_1^i), \Lambda_2^i(m_2^i)\bigg) \nonumber\\
        = & \prod_i^N \sum_{\alpha,\beta} 
        \pi(m_1^{i,\alpha}, m_2^{i,\beta})
        {\cal L}_\kappa^\text{int}\bigg(d_i\bigg|m^{i,\alpha}_1, m^{i,\beta}_2, 
        \nonumber\\
        & \Lambda_1^i(m_1^{i,\alpha}|\Upsilon),
        \Lambda_2^i(m_2^{i,\beta}|\Upsilon)\bigg)
        \Delta m^2 .
    \end{align}
    Here, the sum over $\alpha,\beta$ run over 200 primary and secondary mass bins with width $\Delta m$.
    For each bin, $\Lambda$ is determined by the equation of state parameters $\Upsilon$.
\end{enumerate}
\begin{table}[t!]
    \begin{ruledtabular}
    \begin{tabular}{ccccc}
    Parameter             &Unit         & Prior     & Minimum & Maximum \\
    \hline
    $\mathcal{M}$         &$M_\odot$    & Uniform   & 1.05    & 1.45    \\
    $q$                   &-            & Uniform   & 0.7     & 1       \\
    RA                    &rad.         & Uniform   & 0       & $2\pi$  \\
    DEC                   &rad.         & Cos       & $-\pi/2$& $\pi/2$ \\
    $\cos(\theta_{jn})$   &-            & Uniform   & -1      & 1       \\
    $\psi$                &rad.         & Uniform   & 0       & $\pi$   \\
    $\phi$                &rad.         & Uniform   & 0       & $2\pi$  \\
    $d_L$                 &Mpc          & Comoving  & 10      & 400     \\
    \end{tabular}
    \end{ruledtabular}
    \caption{
    Prior distributions used in our injection set. The intrinsic parameters are the chirp mass $\M$, and mass ratio $q$. The dimensionless tidal deformabilities $\Lambda$ are determined using the SLy equation of state.
    The extrinsic parameters are the right ascension (RA) and declination (DEC), inclination angle $\theta_{jn}$, phase of coalescence $\phi$, polarization angle $\psi$, and luminosity distance $d_L$.
    The comoving prior in the luminosity distance means that we assume a uniform prior in comoving volume. We assume non-spinning binaries in all the injections.
    }
    \label{tab:priors}
\end{table}

\subsection{Interpolation with random forest}\label{sec:Interpolation}
In order to interpolate between samples, we employ the random forest regressor from the {\sc python} package {\sc scikit-learn}~\cite{scikit-learn}.
Broadly speaking, the random forest algorithm combines information from different decision trees. 
A decision tree can be understood as a tree-like model of branches where all outcomes are made based on conditional control statements.
The results of random forest regressors are obtained by averaging over different decision trees~\cite{Breiman:2001,Biau2016}.

The evaluation time and accuracy of the interpolation depends on the number of decision trees and the number of interpolation points.
We use $9\times10^3$ interpolation points, where \unit[85]{\%} of our dataset is used for training and \unit[15]{\%} is used for testing. We find that with 50 decision trees, the interpolation error is $\sim 0.3$\%  and the interpolated-likelihood evaluation time for one value of $\omega=(\M,q,\Lambda_1,\Lambda_2)$ is $\unit[\approx 0.05]{ms}$. Increasing the number of random trees may increase the accuracy, but this level of accuracy is sufficient for our present purposes and the interpolated likelihood evaluation time increases proportionally to the number of decision trees.

\begin{figure}[!t]
    \centering
    \includegraphics[width=\columnwidth]{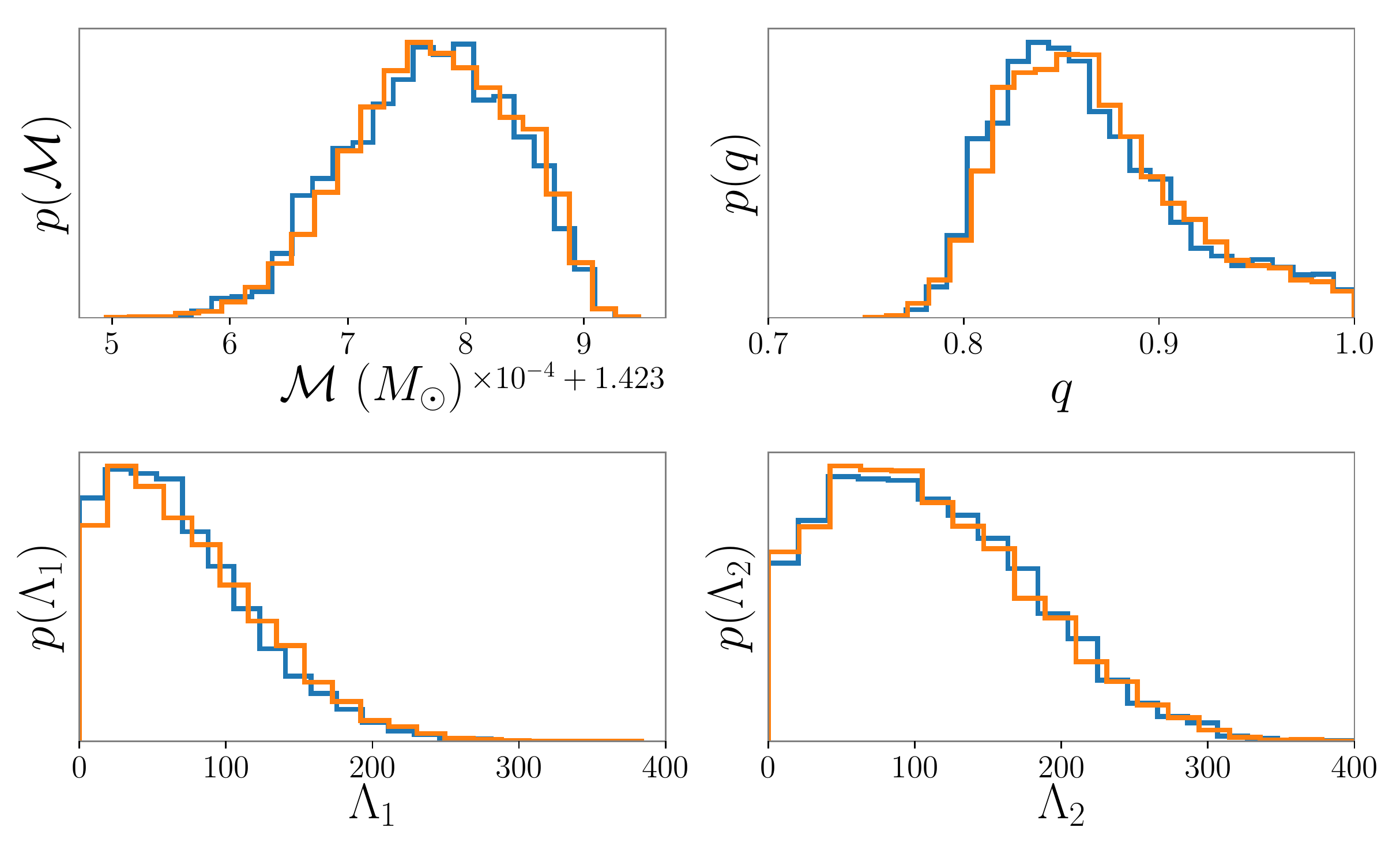}
    \caption{
    Posterior distributions of the intrinsic parameters $\omega=(\M,q,\Lambda_1,\Lambda_2)$. The orange posteriors are obtained by sampling the interpolated likelihood distribution. Our results are compared to the original posteriors shown in blue. This plot indicates than an interpolation error of $\sim 0.3 $\% reproduces the original posteriors. 
    }
    \label{fig:interpolated_posteriors}
\end{figure}

In order to vet the interpolation, we sample the new interpolated-likelihood to obtain posterior distributions of the intrinsic parameters. The posterior distributions from the exact likelihood and the interpolated likelihood are shown in Fig.~\ref{fig:interpolated_posteriors}. The interpolated likelihood distributions shown in Fig.~\ref{fig:interpolated_posteriors} are consistent with the exact likelihood, which indicates that the random forest interpolation is effective.

There are other machine-learning algorithms that could be as effective as random forests, for example, neural network regressors~\cite{tensorflow2015-whitepaper,deep_learning:2015}. While these algorithms have been used in recent studies, exploring the advantages/disadvantages of these algorithms is left for future study.
All our simulations are done within the Bayesian inference library \bilby~\cite{Ashton:2019} and the LIGO algorithm library \lalsuite~\cite{lalsuite}. 
We use the {\sc Dynesty} nested-sampler~\cite{dynesty}.

\section{Mock data study}\label{sec:Results}
We demonstrate our method with a mock data study.
We assume a detector network consisting of the two Advanced LIGO detectors and Advanced Virgo.
We assume that LIGO and Virgo operate at the expected sensitivity of the third observation run (O3) for one year, plus eight months of design sensitivity.
We assume that the binary neutron star range of LIGO Hanford and LIGO Livingston at design sensitivity is \unit[190]{Mpc}, while the range of Virgo at design sensitivity is \unit[140]{Mpc}.
During O3, we additionally assume that the binary neutron star range of LIGO Hanford, LIGO Livingston and Virgo are \unit[100]{Mpc}, \unit[130]{Mpc} and \unit[50]{Mpc} respectively~\cite{observing_scenarios:2018}.

We assume a merger rate of \unit[1540]{Gpc$^{-3}$yr$^{-1}$}, which is consistent with a realistic binary neutron star merger rate~\cite{LSC_GW170817}.
Based on this rate, we expect to detect $N=40^{+10}_{-11}$ binary neutron star mergers (90\% credible interval) with a network matched filter signal-to-noise ratio exceeding the usual threshold of 12.
In Tab.~\ref{tab:priors} we show the prior distributions of our injection set. This results in binaries with individual masses between \unit[1]{$M_\odot$} and \unit[2]{$M_\odot$}.
The tidal deformabilities are determined using the SLy equation of state~\cite{Douchin:2001sv}.

Due to the fact that binary neutron stars can be in-band for long periods of time---for example GW170817 was in band for $\unit[\approx 2]{minutes}$~\cite{LSC_GW170817}--- we use a reduced-order-model (ROM) gravitational waveform to accelerate Bayesian parameter estimation~\cite{Smith:2016}. 
A reduced-order model removes redundant waveform calculations by reduced-order quadrature (ROQ) integration. 
In this study, we use an ROQ implementation of the {\sc IMRPhenomD\_NRTidal} waveform approximant~\cite{Husa:2016,Dietrich:2017,Khan:2016}, which is a spin-aligned waveform that includes tidal interactions and which models the inspiral, merger and ringdown. This allows us to run full parameter estimation for each gravitational-wave measurement in \unit[$\approx 4$]{hours}. 
This is an improvement compared to Ref.~\cite{Lackey:2015}, where they used the {\sc TaylorF2}, {\sc TaylorT1} and {\sc TaylorT4}  approximants without an ROQ implementation of these waveforms. 

\begin{figure}[!t]
    \centering
    \includegraphics[width=\columnwidth]{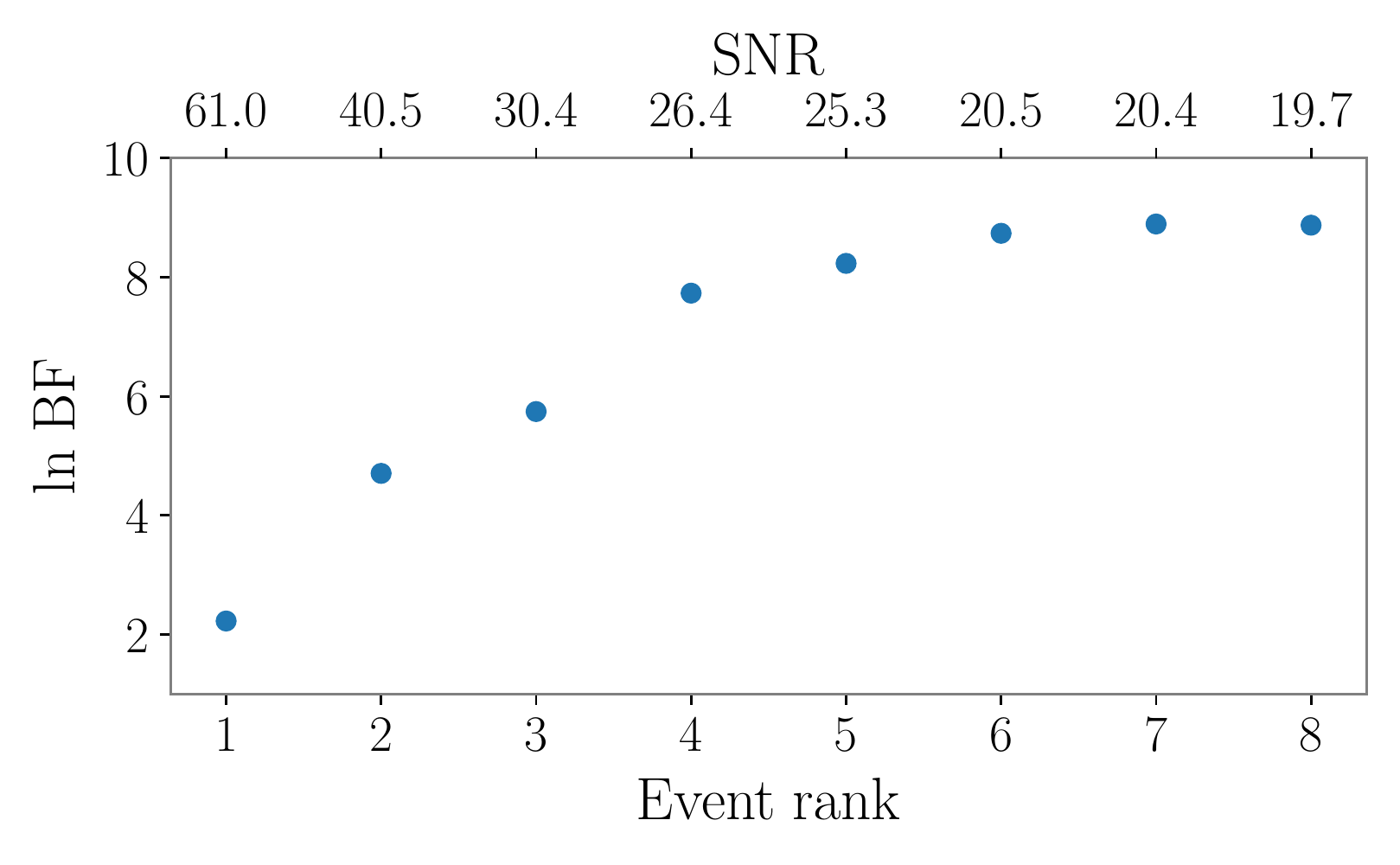}
    \caption{
    Cumulative $\ln$(BF) as a function of event rank. We rank events from loud to quiet, where event rank=1 is defined as the loudest event. The ln BF is calculated between the hypothesis that $\Lambda \neq 0$ and $\Lambda=0$. The event rank is shown in the lower horizontal axis and its corresponding SNR is shown in the upper horizontal axis. We find that after $\text{SNR} \approx 20$, the ln BF starts to plateau. This means that past $\text{SNR}\approx 20$, signals do not significantly add substantial information about the equation of state. Our analysis focuses only on signals where $\text{SNR}>20$.  
    }
    \label{fig:log_bf}
\end{figure}

We investigate the result from \citet{Lackey:2015}, which states that most of the equation of state information comes from the loudest \unit[$\sim 12$]{\%} of events.
We calculate the Bayes factor comparing the non-black hole  evidence $\mathcal{Z}_{\Lambda \neq 0}$ to the black hole evidence $\mathcal{Z}_{\Lambda=0}$,
\begin{equation}\label{eq:log_bf}
    \text{BF} = \frac{\mathcal{Z}_{\Lambda \neq 0}}{\mathcal{Z}_{\Lambda=0}} .
\end{equation}
Then, we evaluate (\ref{eq:log_bf}) for events with a range of different signal-to-noise ratios. We rank events from loud to quiet, where ``event rank''=1 is defined as the loudest event. Our results are shown in Fig.~\ref{fig:log_bf}, where we plot the cumulative $\ln \text{BF}$ as a function of ``event rank''.
We see that the $\ln{\text{BF}}$ plateaus at $\text{SNR} \approx 20$. This means that events with $\text{SNR}<20$ do not add significant information about the equation of state, compared to the loudest events in our dataset. We find that most of the information comes from the loudest \unit[$\sim 15$]{\%} of events, which is consistent with Ref.~\cite{Lackey:2015}. The remainder of our study focuses only on analyzing the loudest eight events. 

\begin{figure}[!t]
    \centering
    \includegraphics[width=\columnwidth]{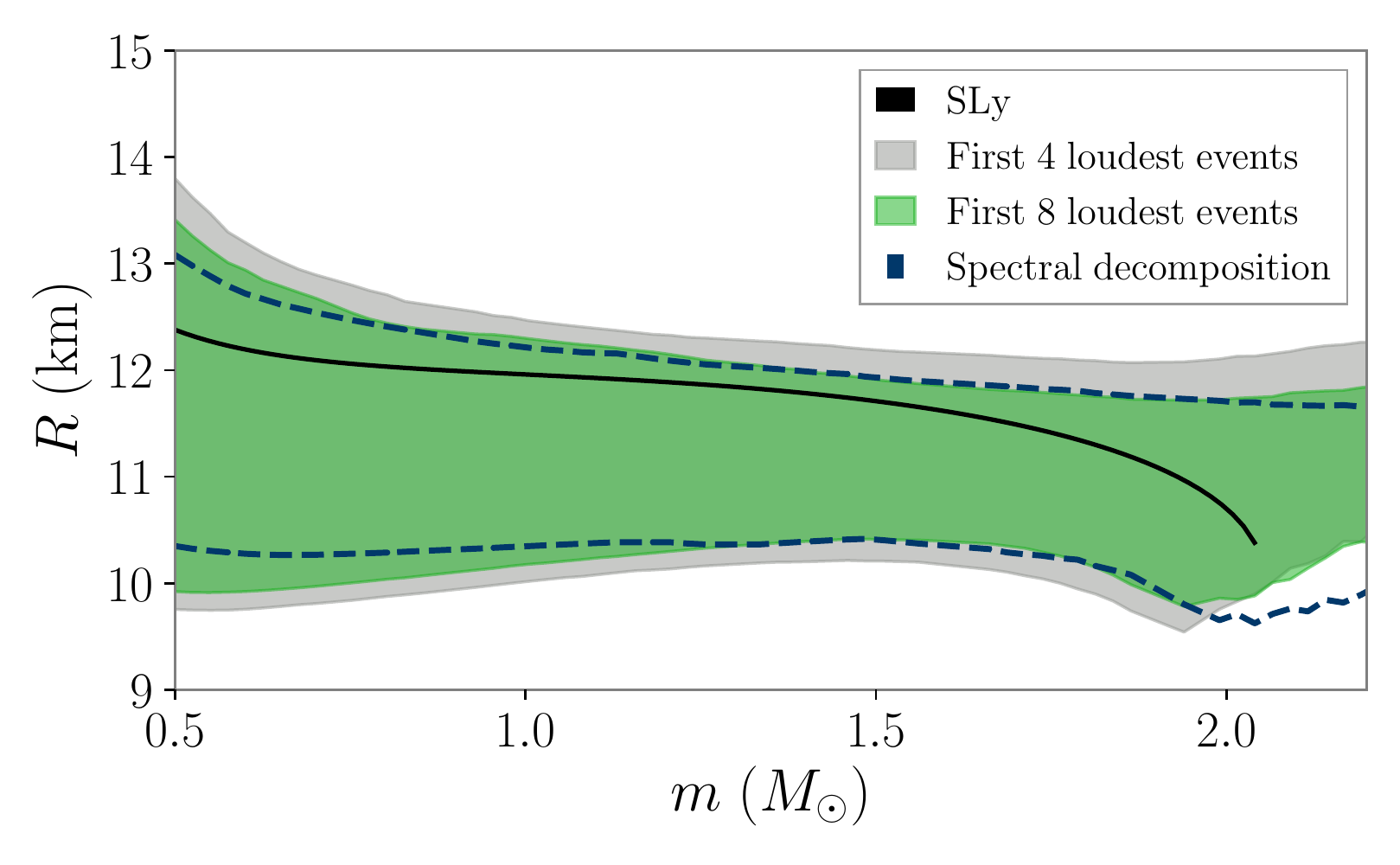}
    \includegraphics[width=\columnwidth]{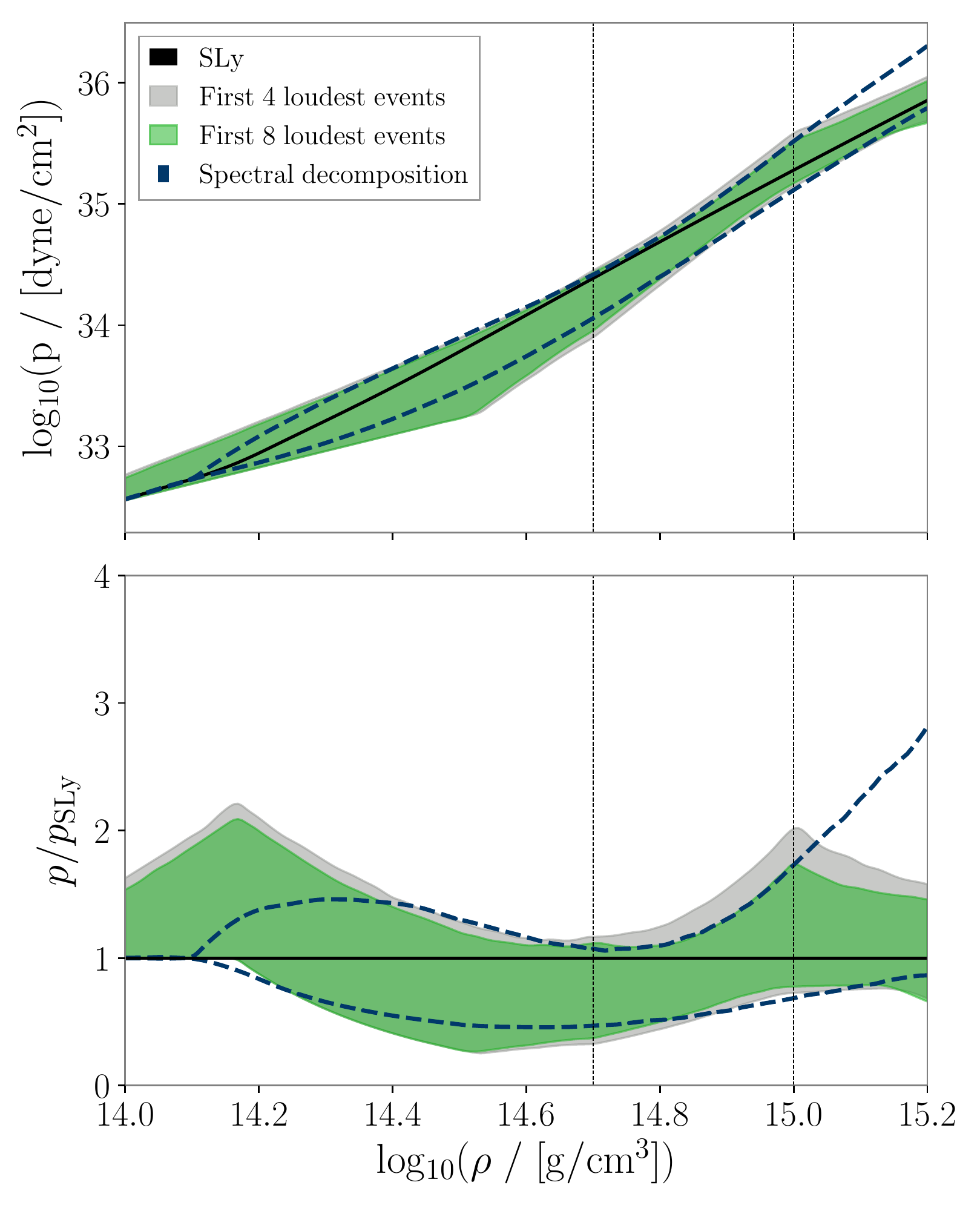}
    \caption{
        Top panel: 90\% credible interval of the posterior distribution of radius as a function of mass. The shaded grey and green posteriors are obtained from analyzing respectively the loudest four and eight events from our dataset with the piecewise polytrope parametrization. The dark-blue dashed posteriors are obtained from analyzing the loudest eight events of our dataset with the spectral decomposition parametrization. 
        Bottom panel: 90\% credible interval of the pressure as a function of radius and the ratio between the pressure and the pressure given by the SLy equation of state $p_{\text{SLy}}$. The gray vertical lines represent the dividing densities of the piecewise polytrope. We assume that binary neutron stars follow the SLy equation of state and the merger rate is fixed to \unit[1540]{Gpc$^{-3}$ yr$^{-1}$}.
    }
    \label{fig:posteriors}
\end{figure}

In Fig.~\ref{fig:posteriors} we show the joint posterior for $(R, m)$ and $(p,\rho)$ obtained using the loudest eight events in our mock data study.
Assuming the piecewise polytrope parametrization of the equation of state, we find that the radius of a \unit[1.4]{$M_\odot$} neutron star is constrained to better than 10\% at 90\% confidence corresponding to a precision of \unit[$\pm 1$]{km}.
At the same time, the pressure at twice the nuclear saturation density can be constrained to better than 45\% at 90\% confidence. 

In order to explore how the constraint on the neutron star equation of state depends on the equation of state parametrization, we analyze the same dataset with the spectral decomposition parametrization~\cite{Lindblom:2010}. Using the same priors outlined in Ref.~\cite{Carney:2018}, we find that the radius of a \unit[1.4]{$M_\odot$} neutron star can be constrained to better than \unit[$\sim 10$]{\%}, consistent with the piecewise polytrope parametrization.

\section{Discussion}\label{sec:Discussion}
We show that after an observation time of one year of LIGO/Virgo operating at O3 sensitivity, plus eight months of design sensitivity, we can constrain the neutron star radius to $\sim 10$\% at 90\% confidence. Our results are consistent with Ref.~\cite{Lackey:2015}, where they showed that the neutron star radius can be measured to $\sim 10$\% after one year of LIGO/Virgo design sensitivity. Our results are consistent with Ref.~\cite{Lackey:2015} because the number of detections used in their study correspond roughly to the number of detections used in this study. 

The work in~\cite{Lackey:2015} showed that the vast majority of information comes from the loudest five events. We confirm this statement by calculating the Bayes factor between $\mathcal{Z}_{\Lambda \neq 0}/\mathcal{Z}_{\Lambda = 0}$. From Fig~\ref{fig:log_bf}, we can see that the cumulative Bayes factor significantly grows for loud events, but plateaus for events with SNR$\approx 20$, which confirms the statement that the vast majority of information comes from the loudest few events. Our Bayes factor estimation is possible because we use a nested sampler algorithm, which returns a Bayesian evidence for every parameter estimation run.

In this study, we use an ROQ implementation of the {\sc IMRPhenomD\_NRTidal} waveform approximant~\cite{Husa:2016,Dietrich:2017,Khan:2016}, which includes inspiral, merger and ringdown.
This is in contrast to the waveforms used in Ref.~\cite{Lackey:2015}, which used inspiral-only waveforms containing post-Newtonian (PN) terms up to 3.5PN, in the point-particle terms. Their study shows that not including 4PN terms adds systematic biases which can sometimes be larger than the 95\% credible interval of their mass and pressure constraints. In this study, we use the {\sc IMRPhenomD\_NRTidal} approximant, which is calibrated with effective-one-body and numerical relativity waveforms.

\citet{Agathos:2015} showed that a poorly chosen prior for neutron-star mass can bias inferences about the neutron-star equation of state.
While we do not account for prior mismatch in our study, this effect should be taken into account when data from many binary neutron star detections are combined.
The long-term solution, we believe, is to employ a population model, which can be fit using hierarchical modeling~\cite{Thrane:2018qnx}.
Such models are already used to infer the shape of the binary black hole mass spectrum~\cite{o2_pop}.
In this way, the neutron-star mass prior for each event is informed by the distribution inferred by the other events.

In this study, we explore in greater detail a different method to interpolate the likelihood distribution using a random forest regressor which leads to an interpolation error of $\sim 0.3$\%. Our method is different from the Gaussian kernel density estimation algorithm used in~\cite{Lackey:2015}. The interpolation method is important because a poor interpolation accuracy can bias the hyper-posterior $p(\Upsilon|d)$. 

One factor that could change our results is the choice of priors for the piecewise polytrope parametrization. Particularly,~\cite{Carney:2018} shows that different low-density starting points of the piecewise polytrope parametrization result in different mass-radius constraints. In this study, we followed the parametrization used in Refs.~\cite{Read:2009,Lackey:2015}. 

In our analysis, we require that the equation of state allow for a maximum mass in excess of \unit[$m=1.97$]{$M_\odot$} in order to be consistent with the pulsars PSR J0348+0432 and PSR J0740+6620. However, Miller et al.~\cite{Miller:2019nzo} argue that a smooth prior on maximum mass (informed by radio observations) is preferable to a hard cut-off. In future work, it will be useful to revisit prior constraints on the maximum neutron star mass.

We compare our results using two different parametrizations of the equation of state: the piecewise polytrope and the spectral decomposition parametrization. Although these parametrizations are consistent between \unit[$ \rho\approx10^{14.2}$]{g cm$^{-3}$} and \unit[$\rho\approx10^{15}$]{g cm$^{-3}$},  we note differences at densities near \unit[$10^{14.1}$]{g cm$^{-3}$} and \unit[$\gtrsim 10^{15}$]{g cm$^{-3}$}.  The differences near \unit[$10^{14.1}$]{g cm$^{-3}$} are explained by the choice of reference pressure $p_0$,  where we use the value of \unit[$p_0=5.36 \times10^{32}$]{dyne cm$^{-2}$} to reflect that the equation of state is well constrained below the nuclear density~\cite{Carney:2018}. 

We find that the differences above \unit[$\approx 10^{15}$]{g cm$^{-3}$} are  explained by the fact that this region of parameter space is mostly dominated by the prior on causality and high mass neutron star observations. Since the prior that we use in the piecewise polytrope is narrower above \unit[$\approx 10^{15}$]{g cm$^{-3}$}, we find that the piecewise polytrope is more tightly constrained above this density. In particular, we find that the pressure at six times the nuclear saturation density is constrained to $\sim 30$\% and $\sim 50$\% assuming the piecewise polytrope and spectral decomposition parametrization, respectively. 

Another algorithm that can be used to infer equation of state hyper-parameters is {\sc RIFT}~\cite{Lange:2018pyp}. This method interpolates the likelihood distribution over a grid of source parameters using a Gaussian process regressor. In order to recover the parameters of astrophysical sources, the algorithm obtains an interpolated continuous representation of the likelihood, which can also be used to overcome the ``stacking problem'' that we address here in our study.

Finally, in some circumstances, it is possible that we will observe bias in our hyper-posterior $p(\Upsilon|d)$ due to small errors in our interpolation. If this happens, we can fix the problem with an iterative procedure. We can carry out steps 1-5 of our interpolation method described in Sec.~\ref{subsec:Formalism} in order to obtain a posterior predictive distribution in $\omega$. We can use the posterior predictive distribution to choose additional interpolation points $\omega_j$. This amounts to an adaptive mesh refinement where the interpolated likelihood $\mathcal{L}_k$ is calculated with greater density in $\omega$ depending on the data, helping to control systematic error as we combine more events. This ability to adaptively refine the interpolation is a potential advantage over Gaussian kernel density estimation.

\section{Conclusion}\label{sec:Conclusion}
We present a new method to combine information from an ensemble of gravitational-wave observations to constrain the neutron-star equation of state. Our method interpolates the likelihood distribution with an error of 0.3\% using a random forest regressor. Our interpolation algorithm is an alternative to the Gaussian kernel density estimation method used in~\cite{Lackey:2015}.
We demonstrate how LIGO and Virgo are likely to constrain the radius and pressure of binary neutron stars with the first forty detections. We show that the radius of a \unit[1.4]{$M_\odot$} neutron star can be constrained to $\sim 10$\% at 90\% confidence and the density at twice the nuclear saturation density can be constrained to $\sim 45$\% at 90\% confidence.
Finally, we combine gravitational-wave measurements by running full parameter estimation on events with SNR$\geq 20$ using an ROQ implementation of the {\sc IMRPhenomD\_NRTidal} approximant, which significantly reduces the cost of parameter estimation. We add Appendix A following the publication of the manuscript in the journal where we calculate a posterior distribution for the maximum, non-rotating neutron star mass.  The Appendix does not appear in the published journal version of the paper, but only in the arXiv version.

\section*{Acknowledgments}
This work is supported through Australian Research Council Grant No. CE170100004, No. FT150100281, No. FT160100112, and No. DP180103155.
F.H.V. is supported through the Monash Graduate Scholarship (MGS).
The results presented in this manuscript were calculated using the computer clusters at California Institute of Technology and Swinburne University of Technology (OzSTAR). The authors are grateful for computational resources provided by the LIGO Laboratory and supported by National Science Foundation Grants No. PHY-0757058 and No. PHY-0823459.
This is LIGO Document No. P1900262.

\appendix
\section{Maximum mass}
\begin{figure}[!t]
    \centering
    \includegraphics[width=\columnwidth]{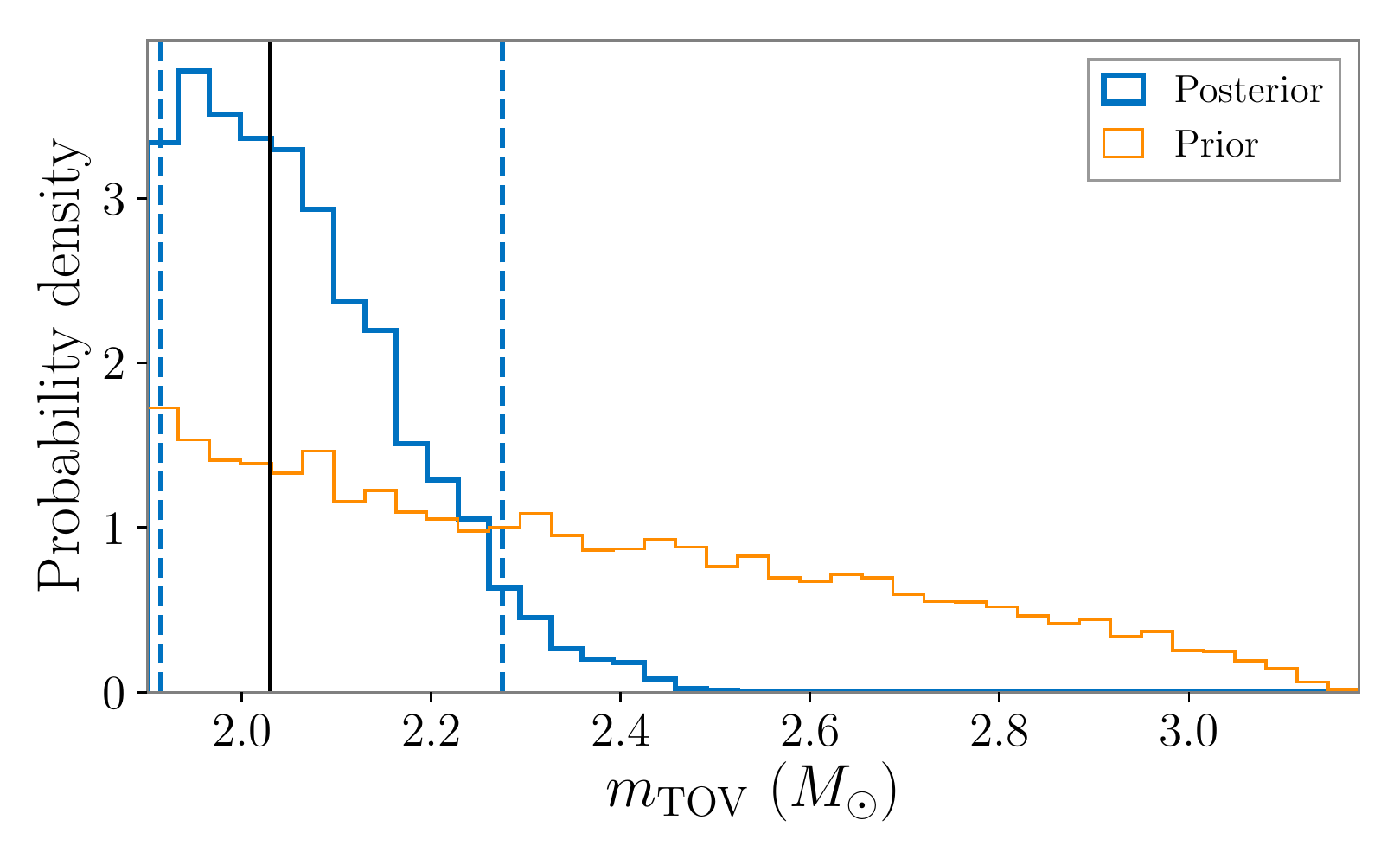}
    \caption{
    Maximum mass posterior distribution using the samples from Fig.~\ref{fig:posteriors} assuming the piecewise polytrope parametrization of the equation of state. The priors defined in Tab.~\ref{tab:hyper_priors} imply a prior on the maximum neutron star mass which is shown in orange. The true value is represented by the vertical black line and the 90\% confidence intervals are shown as blue dotted lines. We find that the TOV mass can be constrained with an accuracy of \unit[$\sim 0.3$]{$M_\odot$} (90\% confidence).
    }
    \label{fig:tov_mass}
\end{figure}

The posterior distributions shown in Fig.~\ref{fig:posteriors} imply a posterior on the maximum neutron star mass allowed by the equation of state (TOV mass). Assuming the piecewise polytrope parametrization and the priors defined in Tab.~\ref{tab:hyper_priors}, we calculate the TOV mass posterior distribution using the samples from Fig.~\ref{fig:posteriors}. Note that we reject all samples with maximum masses  below \unit[1.9]{$M_\odot$}, conservatively consistent with the maximum masses observed from PSR J0348+0432 ~\cite{Antoniadis1233232} and PSR J0740+6620~\cite{Cromartie:2019}. The result is shown in Fig.~\ref{fig:tov_mass}, where we find that the TOV mass can be measured with an accuracy of \unit[$\sim 0.3$]{$M_\odot$} (90\% confidence). This result will differ for different assumed equations of state; we leave this exploration to future work.

\bibliography{bibliography}

\end{document}